\documentclass{jpsj3}
\usepackage{txfonts}
\usepackage{color}

\title{Competition between Interactions and Randomness in Photoinduced Synchronization of Charge Oscillations on a Dimer Lattice}

\author{Toshiya Shimada and Kenji Yonemitsu\thanks{E-mail: kxy@phys.chuo-u.ac.jp}}
\inst{Department of Physics, Chuo University, Bunkyo, Tokyo 112-8551, Japan} %\\
\abst{ 
The synchronization of charge oscillations after photoexcitation that has been realized through the emergence of an electronic breathing mode on dimer lattices is studied here from the viewpoint of the competition between interactions and randomness. We employ an extended Hubbard model at three-quarter filling on a simple dimer lattice and add random numbers to all transfer integrals between nearest-neighbor sites. Photoinduced dynamics are calculated using the time-dependent Schr\"odinger equation by the exact diagonalization method. Although the randomness tends to unsynchronize charge oscillations on different bonds during and after photoexcitation, sufficiently strong on-site repulsion $U$ overcomes this effect and synchronizes these charge oscillations some time after strong photoexcitation. The degree of synchronization is evaluated using an order parameter that is derived from the time profiles of the current densities on all bonds. As to the nearest-neighbor interaction $V$, if $V$ is weakly attractive, it increases the order parameter by facilitating the charge oscillations. The relevance of these findings to previously reported experimental and theoretical results for the organic conductor $\kappa$-(bis[ethylenedithio]tetrathiafulvalene)$_2$Cu[N(CN)$_2$]Br is discussed. 
 }

\begin{document}
\maketitle

\section{Introduction}
Photoinduced phase transitions are nonlinear dynamical phenomena, in which events on different timescales are involved.\cite{koshigono_jpsj06,yonemitsu_pr08,basov_rmp11,nicoletti_aop16,giannetti_aip16,kaiser_ps17,ishihara_jpsj19} The transient lowering of the symmetry of a many-electron state is an important issue, and qualitative progress in its observation and understanding has been made by developments in experimental techniques. 
In equilibrium and continuous phase transitions accompanying symmetry breaking, a long-range order is formed by the spontaneous development of fluctuations. In photoinduced phase transitions, a conventional picture is similar, where fluctuations are produced by photoexcitation.\cite{nagaosa_prb89} The development of fluctuations is achieved by interactions; thus, photoinduced phase transitions are cooperative phenomena. Their stochastic processes are described with a probability. 

As the pulse width becomes small and the amplitude of the optical field becomes large, photoinduced dynamics can be qualitatively changed. The application of an intense optical field transiently and directly lowers the symmetry of a many-electron state, keeping the coherence in many-electron motion\cite{kawakami_prl10,matsubara_prb14} and leading to transient charge order formation before relaxation becomes significant.\cite{yonemitsu_jpsj18a,nag_prb19} Such an order is absent before photoexcitation, and it would oscillate and become zero on average. 

The excited states that are responsible for ultrafast dynamics inevitably have high energies. The number of such states is large. Dephasing of charge oscillations is often significant especially when electron correlations are strong. In this context, it is nontrivial for electrons to oscillate coherently. In any case, a coherent charge oscillation with a short period is important for the ultrafast lowering of the symmetry. To reduce the effect of dephasing, it is advantageous for charge oscillations to be synchronized. Here, we study competing effects in the photoinduced synchronization of charge oscillations that are previously reported on dimer lattices.\cite{yonemitsu_jpsj18a,yonemitsu_jpsj18b}

In the organic superconductor $\kappa$-(bis[ethylenedithio]tetrathiafulvalene)$_2$Cu[N(CN)$_2$]Br [$\kappa$-(BEDT-TTF)$_2$Cu[N(CN)$_2$]Br] with a dimerized structure, a nonlinear charge oscillation and a resultant stimulated emission have been observed on the high-energy side of the main reflectivity spectrum only after strong photoexcitation.\cite{kawakami_np18} Theoretically, by using an extended Hubbard model at three-quarter filling on a dimer lattice that corresponds to this compound\cite{yonemitsu_jpsj18a} and other models on dimer lattices,\cite{yonemitsu_jpsj18a,yonemitsu_jpsj18b} we can realize a nonlinear charge oscillation only after strong photoexcitation that is characterized as an electronic breathing mode. 

It has already been pointed out that sufficiently strong on-site repulsion is necessary for this mode to be dominant over any charge oscillations appearing in the linear conductivity spectra. The synchronization of charge oscillations between charge-rich and charge-poor sites with the help of this interaction is suggested by the fact that time-averaged bond charge densities on different bonds governed by different transfer integrals are similar functions of the amplitude of the optical field (before taking the time average) when the on-site repulsion is sufficiently strong.\cite{yonemitsu_jpsj18a} 

In this paper, we add random numbers to transfer integrals to intentionally weaken the synchronization of charge oscillations. A synchronization order parameter is defined\cite{kuramoto_book84,acebron_rmp05,rodrigues_pr16,nag_prb19} using the phases of the oscillating current densities on all bonds connected by transfer integrals. Then, we investigate the competition between interactions and random transfer integrals. Thus, we directly show that the emergence of an electronic breathing mode is caused by the synchronization of charge oscillations through the on-site repulsion. 

\section{Dimerized Model with Disorder in Two Dimensions}
\begin{figure}
\includegraphics[height=9.0cm]{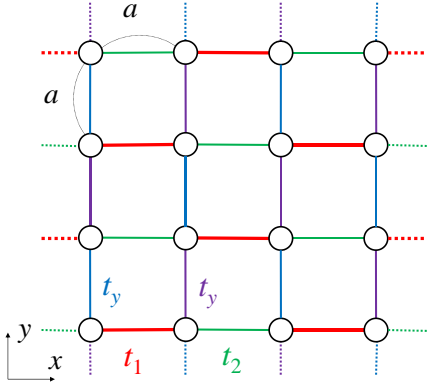}
\caption{(Color online) 
Two-dimensional lattice consisting of dimers. The magnitude of the intradimer transfer integral $t_1$ is larger than those of the interdimer transfer integrals $t_2$ and $t_y$. The distance between neighboring sites is $a$ along the $x$- and $y$-axes. 
\label{fig:dimer_latt}}
\end{figure}

In previous studies, the electronic breathing mode is analyzed on a two-dimensional lattice with one type\cite{yonemitsu_jpsj18b} or two types\cite{yonemitsu_jpsj18a} of dimers. Here, we use a square lattice that is similar to but simpler than that used in Ref.~\citen{yonemitsu_jpsj18b}, and configure dimers, ``$ t_1 $ bonds,'' as shown in Fig.~\ref{fig:dimer_latt}. 
We employ an extended Hubbard model at three-quarter filling, 
\begin{eqnarray}
H  &=& \sum_{\langle ij \rangle \sigma} t_{ij} 
( c^\dagger_{i\sigma} c_{j\sigma} + c^\dagger_{j\sigma} c_{i\sigma}  )
+U\sum_i \left( n_{i\uparrow} -\frac34 \right)
\left( n_{i\downarrow} -\frac34 \right) \nonumber \\ &&
+V\sum_{\langle ij \rangle} 
\left( n_i -\frac32 \right)
\left( n_j -\frac32 \right)
\;, \label{eq:model}
\end{eqnarray} 
where $ c^\dagger_{i\sigma} $ creates an electron with spin $ \sigma $ at site $ i $, $ n_{i\sigma} $=$ c^\dagger_{i\sigma} c_{i\sigma} $, and $ n_i $=$ \sum_\sigma n_{i\sigma} $. The parameters $ U $ and $ V $ represent the on-site and nearest-neighbor Coulomb repulsion strengths, respectively. The transfer integral is $ t_{ij} = t_1(1+\delta_{ij}) $ inside a dimer along the $x$-axis, $ t_{ij} = t_2(1+\delta_{ij}) $ outside a dimer along the $x$-axis, or $ t_{ij} = t_y(1+\delta_{ij}) $ along the $y$-axis, as shown in Fig.~\ref{fig:dimer_latt}, where $ \delta_{ij} $ are uniformly distributed random numbers on the interval $ [-\epsilon, \epsilon] $. For results shown later, averages are taken over 25 random number distributions unless stated otherwise. A 4$\times$4-site system with periodic boundary conditions is used unless stated otherwise. We use $ t_1 $=$-0.3$, $ t_2 $=$-0.1$, and $ t_y $=$-0.1$. If we regard these values as given in units of eV, they roughly correspond to intradimer and interdimer transfer integrals in dimerized organic conductors $\kappa$-(BEDT-TTF)$_2$X.\cite{mori_bcsj99,watanabe_sm99} In Eq.~(\ref{eq:model}), the constant term is subtracted in such a way that the total energy becomes zero in equilibrium at infinite temperature. 

The initial state is the ground state obtained by the exact diagonalization method. Photoexcitation is introduced through the Peierls phase
\begin{equation}
c_{i\sigma}^\dagger c_{j\sigma} \rightarrow
\exp \left[
\frac{ie}{\hbar c} \mbox{\boldmath $r$}_{ij} \cdot \mbox{\boldmath $A$}(t)
\right] c_{i\sigma}^\dagger c_{j\sigma}
\;, \label{eq:photo_excitation}
\end{equation}
which is substituted into Eq.~(\ref{eq:model}) for each combination of sites $ i $ and $ j $ with relative position $ \mbox{\boldmath $r$}_{ij}=\mbox{\boldmath $r$}_j-\mbox{\boldmath $r$}_i $. Hereafter, we use $e$=$a$=$\hbar$=$c$=1. We employ symmetric one-cycle electric-field pulses\cite{yonemitsu_jpsj18a,yonemitsu_jpsj18b,yonemitsu_jpsj17a,yonemitsu_jpsj15,yanagiya_jpsj15} and use the time-dependent vector potential 
\begin{equation}
\mbox{\boldmath $A$} (t) = \frac{c\mbox{\boldmath $F$}}{\omega_c} \left[ \cos (\omega_c t)-1 \right]
\theta (t) \theta \left( \frac{2\pi}{\omega_c}-t \right)
\;, \label{eq:monocycle_pulse}
\end{equation}
where the central frequency $ \omega_c $ is chosen to be $ \omega_c = 0.7 $ throughout the paper because the qualitative results are independent of its value as in previous studies.\cite{yonemitsu_jpsj18a,yonemitsu_jpsj18b} The electric field is polarized along $ (1, 1) $ and its maximum is $ \mbox{\boldmath $F$} = (F, F) $, although the qualitative results are unaltered by different choices of the polarization. 
The time-dependent Schr\"odinger equation is numerically solved by expanding the exponential evolution operator with a time slice $ dt $=0.02 to the 15th order and by checking the conservation of the norm.\cite{yonemitsu_prb09}

\section{Effect of Randomness on Electronic Breathing Mode}
In our previous paper where we did not consider randomness on dimer lattices,\cite{yonemitsu_jpsj18a} Fourier spectra of the intradimer charge disproportionation after photoexcitation are shown to have characteristics as follows. After weak photoexcitation with small $F$, they have peaks at energies where the conductivity spectrum in the ground state has peaks. After strong photoexcitation with large $F$, they have one dominant peak due to the electronic breathing mode, which is a nonlinear charge oscillation assisted by the on-site repulsion $U$. If the interactions are absent, all of the charge oscillations are linear and do not decay after photoexcitation, so that all of their peaks are high. If the interactions are present, on the other hand, all of the charge oscillations finally decay owing to dephasing. However, the electronic breathing mode for large $F$ decays much slower than the linear charge oscillations appearing in the conductivity spectrum. As a result, the electronic breathing mode becomes dominant for sufficiently strong interaction $U$.\cite{yonemitsu_jpsj18a} The synchronization of charge oscillations is supposed to contribute to the relatively slow decay of the electronic breathing mode, or equivalently, to the suppression of the other charge oscillation modes. The degree of synchronization would be controlled by randomness. Then, we study the effect of randomness introduced into transfer integrals on this mode. 

\begin{figure}
\includegraphics[height=6.2cm]{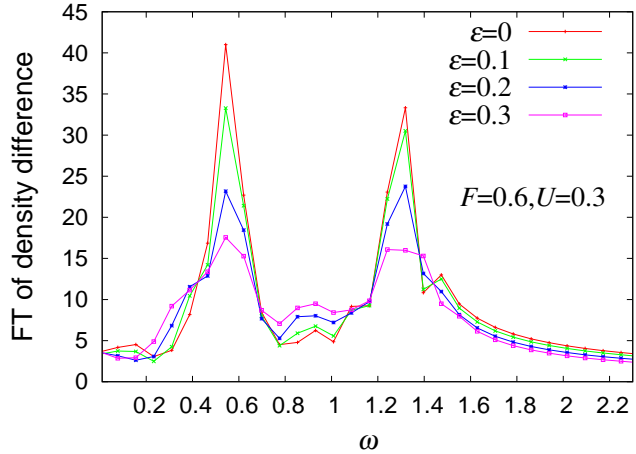}
\caption{(Color online) 
Absolute values of Fourier transforms of random-number-distribution-averaged time profiles ($ T < t < 10T $) of charge-density difference between sublattices, for different values of $\epsilon$, $F$=0.6, $U$=0.3, and $V$=0. 
\label{fig:ftcd_vs_omega_f0p6u0p3v0p0_25seed}}
\end{figure}
From the time profile of the difference between the total charge densities at the left and right sites of the dimers in a time span of $ T < t < 10T $ with $ T=2\pi/\omega_c $, we calculate its Fourier transform for each random number distribution and average the Fourier transforms over random number distributions with a fixed value of $\epsilon$. The absolute values of the transforms are shown in Fig.~\ref{fig:ftcd_vs_omega_f0p6u0p3v0p0_25seed} for large $F$. As in the case without randomness, the electronic breathing mode at $ \omega_{\mbox{osc}} \equiv 2\left( \mid t_1 \mid + \mid t_2 \mid +2 \mid t_y \mid \right) $=1.2\cite{yonemitsu_jpsj18a,yonemitsu_jpsj18b} becomes dominant. In the present dimer lattice of a small size, the peak below 0.6 is also noticeable, but mean-field calculations indicate that this peak becomes less noticeable for larger sizes and merged into a continuum spectrum in the thermodynamic limit. The frequency of the electronic breathing mode is almost independent of the interval $ [-\epsilon, \epsilon] $ of random numbers, indicating that this mode does not lose its identity even if its charge oscillations are inhomogeneous. Later, we will demonstrate that charge oscillations on different bonds are indeed synchronized for sufficiently large $U$ to maintain its identity. 

\section{Definition of Synchronization Order Parameter}
\begin{figure}
\includegraphics[height=13.6cm]{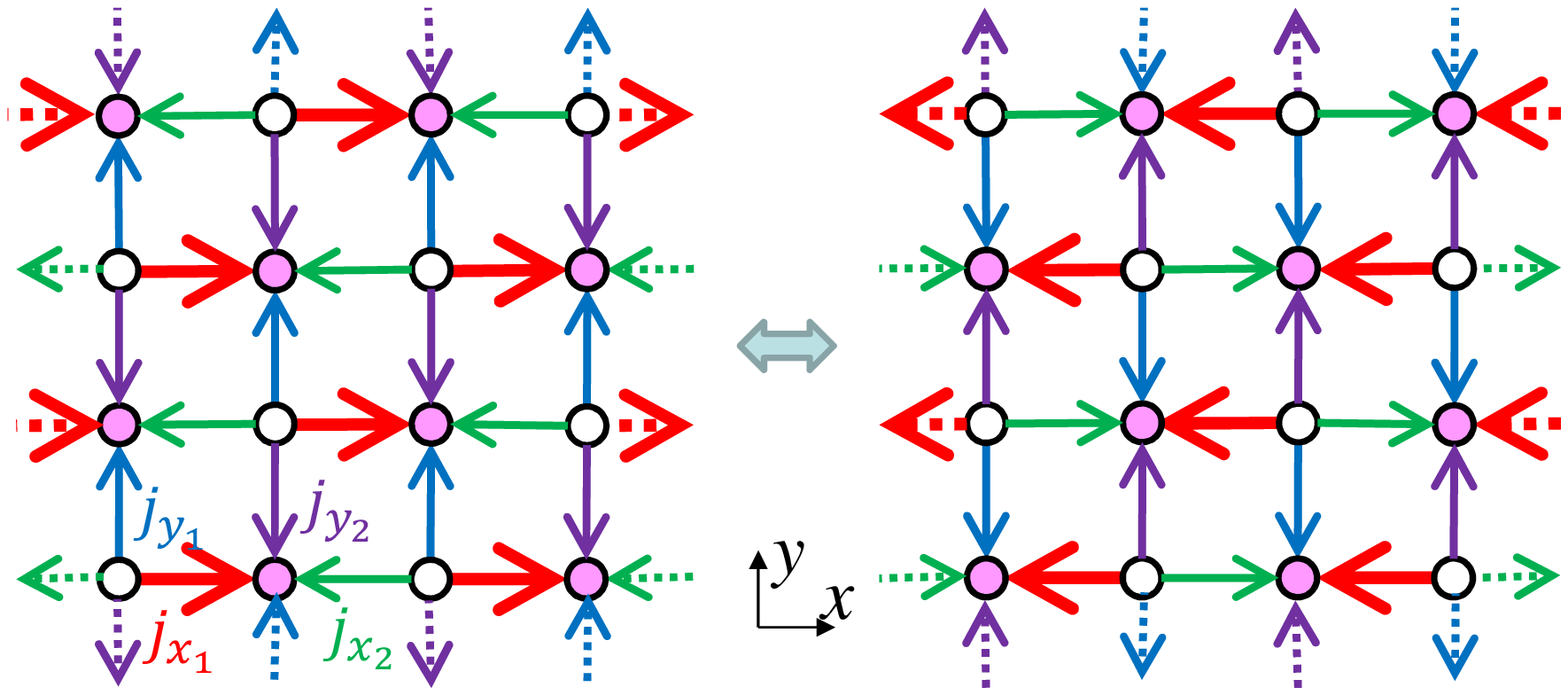}
\caption{(Color online) 
Current distribution in electronic breathing mode. 
\label{fig:breathing}}
\end{figure}

\begin{figure}
\includegraphics[height=16.0cm]{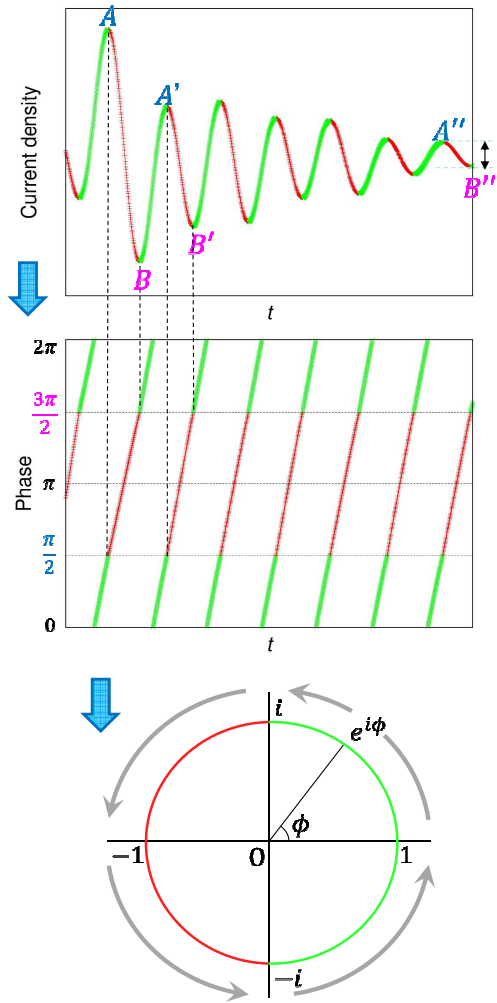}
\caption{(Color online) 
Assignment of phase in defining synchronization order parameter. See text for details. 
\label{fig:order_parameter}}
\end{figure}

To define a measure of how synchronized the charge oscillations are, we refer to the synchronization order parameter that is used in the Kuramoto model.\cite{kuramoto_book84,acebron_rmp05,rodrigues_pr16,nag_prb19} In the electronic breathing mode, the current distribution alternates between the patterns shown in the left and right panels of Fig.~\ref{fig:breathing}. 
Then, we regard the current density on each bond ($ \sum_\sigma \langle -ic^\dagger_{i\sigma} c_{j\sigma} +ic^\dagger_{j\sigma} c_{i\sigma} \rangle $ between sites $i$ and $j$ if $\mbox{\boldmath $A$}(t)$=0) as taking a positive (negative) value if the current flows as in the left (right) panel. We assign the argument $ \phi $ of a complex number $ e^{i\phi} $ to the current density on each bond by following its time evolution as follows. Note that we assign $ \phi $ only when the current density changes in time. When it evolves between local maxima $A$, $A'$, $\cdots$ and local minima $B$, $B'$, $\cdots$ as shown in the upper panel of Fig.~\ref{fig:order_parameter}, we divide the time evolution into intervals $[A,B]$, $[B,A']$, $[A',B']$, and so on. 
When the current density $j_m(t)$ on bond $m$ at time $t$ decreases from $A$ to $B$, we describe its evolution as $ j_m(t) = \frac{A+B}{2}+\frac{A-B}{2}\sin \phi_m(t) $ to obtain $ \phi_m(t) $ in the interval $ \pi/2 \leq \phi_m(t) \leq 3\pi/2 $. When it increases from $B$ to $A'$, we describe its evolution as $ j_m(t) = \frac{A'+B}{2}+\frac{A'-B}{2}\sin \phi_m(t) $ to obtain $ \phi_m(t) $ in the interval $ 3\pi/2 \leq \phi_m(t) \leq 2\pi $ or $ 0 \leq \phi_m(t) \leq \pi/2 $. Thus, we transform the time evolution of $j_m(t)$ into that of $ \phi_m(t) $, as shown in the middle panel of Fig.~\ref{fig:order_parameter}. We regard $ \phi_m(t) $ as the argument of a complex number $e^{i\phi_m(t)}$ of magnitude one, as shown in the lower panel of Fig.~\ref{fig:order_parameter}. Then, we define the synchronization order parameter $ r(t) $ at time $ t $ by $ r(t) e^{i \psi(t)} = \frac{1}{M} \sum_{m=1}^{M} e^{i\phi_m(t)} $, where $ M $ is the total number of bonds. The range of values that $ r(t) $ can take is $[0, 1]$. If $A-B$ or $A'-B$ is smaller than 0.01 for $V$=0 or 0.005 for $V\neq 0$, however, we do not regard the time variation as an oscillation and do not define $ \phi_m(t) $ or $ r(t) $. 

\section{Competition between Interactions and Randomness}

\subsection{Competition with on-site repulsion}
\begin{figure}
\includegraphics[height=15.6cm]{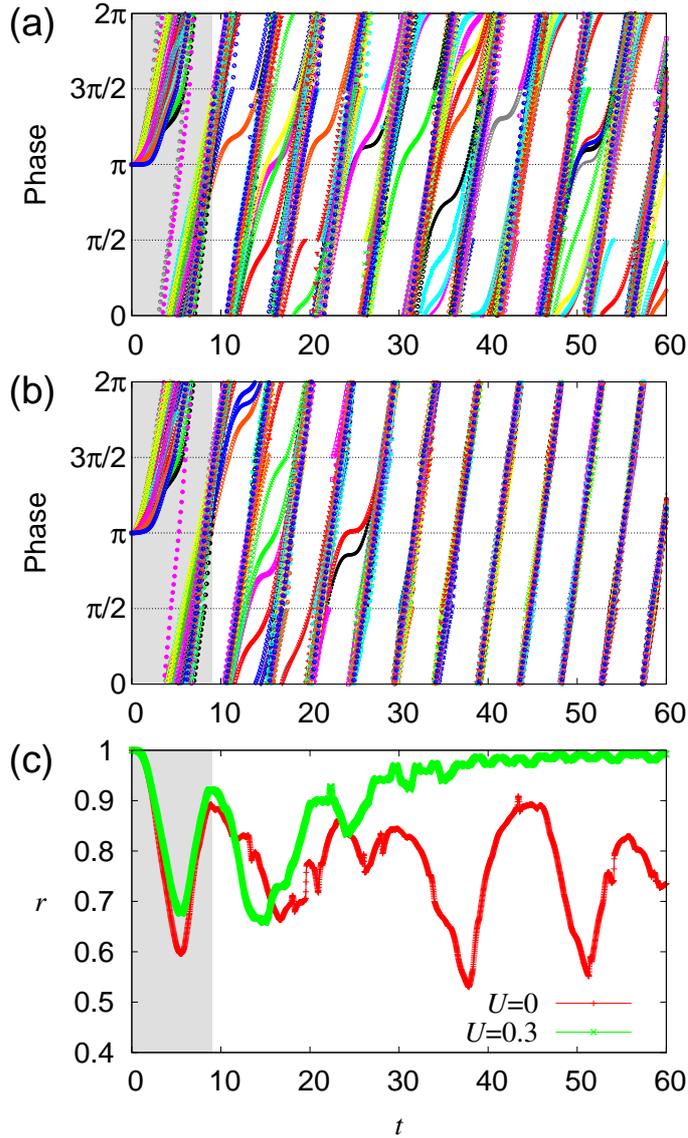}
\caption{(Color online) 
Time profiles of (a) phases for $U$=0, (b) phases for $U$=0.3, and (c) their synchronization order parameters, for $F$=0.6, $V$=0, and a fixed random number distribution with $\epsilon$=0.3. 
\label{fig:phase_op_time}}
\end{figure}
The time profiles of the phases $ \phi_m(t) $ of all bonds $m$ are shown in Figs.~\ref{fig:phase_op_time}(a) and \ref{fig:phase_op_time}(b) for $U$=0 and $U$=0.3, respectively, with $V$=0, a fixed and common random number distribution with $\epsilon$=0.3 as an example, and optical field amplitude $F$=0.6. The duration of photoexcitation is shaded in this figure. In the noninteracting case [Fig.~\ref{fig:phase_op_time}(a)], current densities on different bonds generally oscillate with different phases, although there are some short time intervals where most of the phases take similar values; thus, their behaviors are complex and depend on the random number distribution. However, with sufficiently strong on-site repulsion [Fig.~\ref{fig:phase_op_time}(b)], current densities and consequently charge oscillations are synchronized, although they are not synchronized immediately after photoexcitation; thus, it takes some time to synchronize them. 

This observation is clarified in Fig.~\ref{fig:phase_op_time}(c), which presents the time profiles of the synchronization order parameters calculated from the phases shown in Figs.~\ref{fig:phase_op_time}(a) and \ref{fig:phase_op_time}(b). The plots around $t$=20 after photoexcitation $0<t<T\simeq$9 show that the increase in the order parameter occurs earlier for $U$=0.3 than in the noninteracting case. This tendency of the increase in the order parameter, which is made to occur earlier by $U$($>$0), is general and found for other random number distributions. Thus, some random-number-distribution- and $U$-dependent time after photoexcitation, the repulsive interaction $U$ synchronizes the charge oscillations. Considering that the interdimer distances between neighboring molecules of about 6 \AA \space in $\kappa$-(BEDT-TTF)$_2$X\cite{mori_bcsj99,watanabe_sm99} and an electric field of about 10 MV/cm used in the experiment\cite{kawakami_np18} correspond to 0.6 eV, the present field amplitude of $F$=0.6 is comparable to this 0.6 eV although the lattice structure is quite different from that of $\kappa$-(BEDT-TTF)$_2$X. 

\begin{figure}
\includegraphics[height=12.6cm]{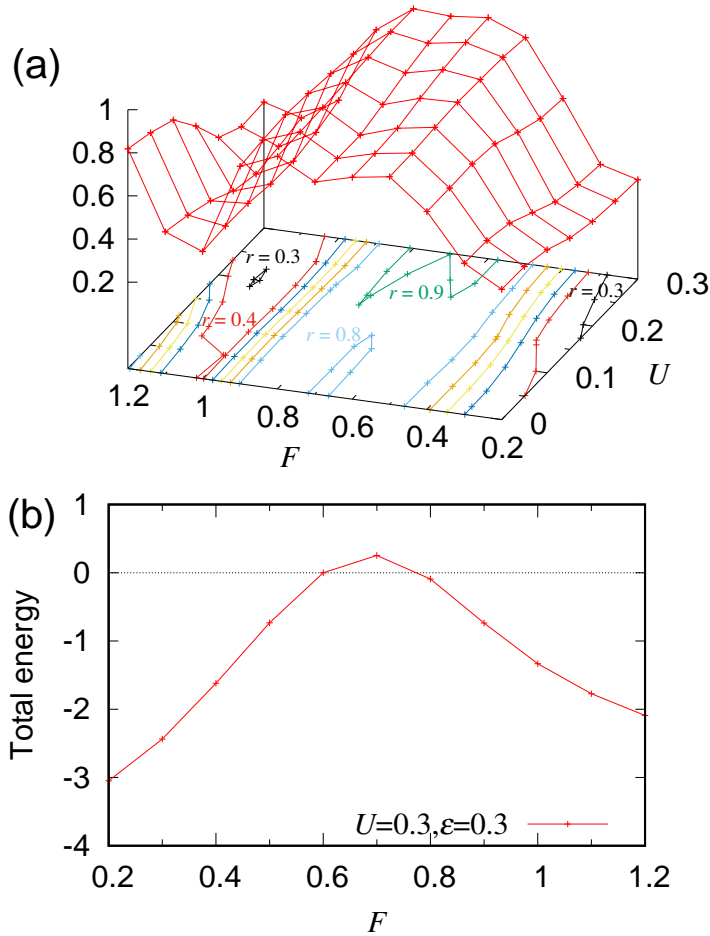}
\caption{(Color online) 
(a) Averaged synchronization order parameter as a function of $F$ and $U$ for $\epsilon$=0.3 and $V$=0. Its contour lines are plotted underneath. (b) Random-number-distribution-averaged total energy after photoexcitation as a function of $F$, for $U$=0.3, $\epsilon$=0.3, and $V$=0. 
\label{fig:op_F_U_Etot_25seed}}
\end{figure}
It is now clear that it takes a nonzero time to synchronize charge oscillations. After a lapse of further time, the charge oscillations decay, the current densities vanish, and we cannot discuss synchronization. Until the current densities vanish, the charge oscillations appear to remain synchronized. Thus, we take the average of the order parameters over a time span of $ 3T<t<6T $ and random number distributions and see how it depends on parameters in the Hamiltonian and the optical field amplitude $F$. In Fig.~\ref{fig:op_F_U_Etot_25seed}(a), we show the averaged synchronization order parameter as a function of $F$ and $U$ for $\epsilon$=0.3 and $V$=0. Its contour lines are plotted underneath. For small $F$, linear charge oscillations with different frequencies contribute to suppressing the order parameter. Around $F$=0.8, the on-site repulsion $U$ suppresses the linear charge oscillations, and the electronic breathing mode becomes relatively dominant; thus, the averaged synchronization order parameter almost reaches the maximum value of 1 for sufficiently large $U$. In the noninteracting case ($U$=0), all charge oscillations are linear and do not decay without dephasing, so that the order parameter is smaller than the repulsive case even when $F$ is large. Therefore, around $F$=0.8, the order parameter is described by an increasing function of $U$. 

For even larger $F$ values, the total energy after photoexcitation becomes smaller than that at $F$=0.7, but it does not reach the ground-state energy at $F$=0, as shown in Fig.~\ref{fig:op_F_U_Etot_25seed}(b). Around $F$=0.7, the total energy goes beyond that in equilibrium at infinite temperature, indicating that it becomes a negative-temperature state with a suppressed rise in the entropy.\cite{tsuji_prl11,tsuji_prb12,yonemitsu_jpsj15,yanagiya_jpsj15} For $F>0.8$, however, the rise in the entropy would become nonnegligible and disturb the synchronization. Similar results are obtained in our previous paper,\cite{yonemitsu_jpsj18a} where the electronic breathing mode becomes less dominant for very large $F$. Thus, the averaged synchronization order parameter is suppressed for such large $F$ values. 

\begin{figure}
\includegraphics[height=11.6cm]{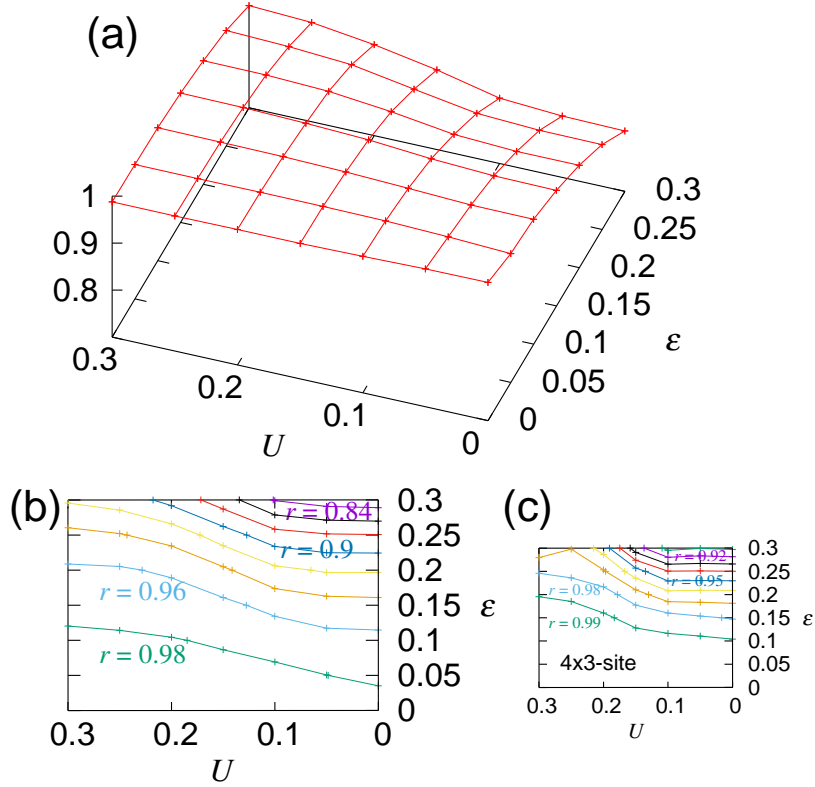}
\caption{(Color online) 
(a) Averaged synchronization order parameter as a function of $U$ and $\epsilon$ for $F$=0.6 and $V$=0. (b) Contour plot of data shown in (a). (c) Similar plot as Fig.~\ref{fig:op_U_epsilon_25seed}(b) for a 4$\times$3-site system with 20 random number distributions. 
\label{fig:op_U_epsilon_25seed}}
\end{figure}
To see the competition between the effect of on-site repulsion $U$ and that of randomness $\epsilon$, we plot the averaged synchronization order parameter in Fig.~\ref{fig:op_U_epsilon_25seed}(a) as their function for $F$=0.6 and $V$=0. Its contour lines are plotted in Fig.~\ref{fig:op_U_epsilon_25seed}(b). Particularly for small $U$, it is apparent that the randomness $\epsilon$ in transfer integrals reduces the order parameter by inhomogeneously modifying the frequencies of charge oscillations. However, for large $U$, $\epsilon$ does not reduce the order parameter very substantially. Even in the case of large $\epsilon$, sufficiently strong on-site repulsion $U$ restores the order parameter. For $U$=$\epsilon$=0, we calculate the order parameter for larger systems and find that it is smaller than that in the present 4$\times$4-site system. The overestimation in the present system is due to the fact that a very small number of charge oscillations, as suggested in Fig.~\ref {fig:ftcd_vs_omega_f0p6u0p3v0p0_25seed}, contribute to the order parameter without decaying. The time span of $ 3T<t<6T $ is too short to distinguish the frequency of the electronic breathing mode from twice the frequency of the low-energy mode. This situation should disappear for larger systems. Therefore, the averaged order parameter is expected to be overestimated in the present small system at least in the small-$U$-small-$\epsilon$ region of Fig.~\ref{fig:op_U_epsilon_25seed}(b). To see such a finite size effect, we calculate the averaged order parameter for an even smaller 4$\times$3-site system with 20 random number distributions and otherwise the same parameters and show it in Fig.~\ref{fig:op_U_epsilon_25seed}(c). The averaged order parameter is indeed overestimated for a smaller system, but the general behavior is essentially the same as that observed in Fig.~\ref{fig:op_U_epsilon_25seed}(b). Thus, we expect that the present finding is valid for larger systems. 

\subsection{Effect of nearest-neighbor interaction}
In our previous paper,\cite{yonemitsu_jpsj18a} we showed that repulsive intersite interactions (with different strengths for different intersite distances on the two-dimensional lattice for $\kappa$-(BEDT-TTF)$_2$Cu[N(CN)$_2$]Br) enhance dephasing and broaden the Fourier spectrum of the charge-density difference. Thus, a repulsive nearest-neighbor interaction $V$ is expected to reduce the synchronization. Here, we see the effect of the nearest-neighbor interaction $V$ in the disordered case of $\epsilon>0$. In reality, $V$ should be positive, but we investigate its effect for both $V>0$ and $V<0$. 

\begin{figure}
\includegraphics[height=11.6cm]{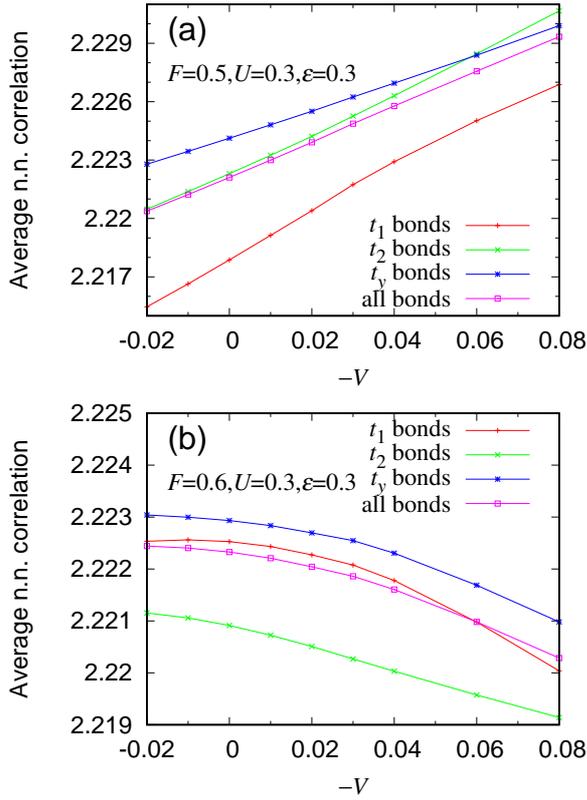}
\caption{(Color online) 
Averaged nearest-neighbor charge-density correlations $ \langle n_i n_j \rangle $ as a function of $-V$ for (a) $F$=0.5 and (b) $F$=0.6, $U$=0.3 and $\epsilon$=0.3. The averages over different types of bonds and those over all bonds are shown. 
\label{fig:n_n_corr_Vm_25seed}}
\end{figure}
It is known in the ground state that a positive $V$ makes charge-density correlations $ \langle n_i n_j \rangle $ for nearest-neighbor sites $i$ and $j$ smaller than those for $V=0$. Even in transient states after photoexcitation with $F\leq0.5$, it decreases $ \langle n_i n_j \rangle $ (i.e., it increases transient charge disproportionation) and decreases the magnitude of the current density on each bond. On the other hand, a negative $V$ makes $ \langle n_i n_j \rangle $ larger and decreases transient charge disproportionation after photoexcitation with $F\leq0.5$. The nearest-neighbor charge-density correlations $ \langle n_i n_j \rangle $ averaged over the time span of $ 3T<t<6T $, random number distributions, and different types of bonds or all bonds are shown in Fig.~\ref{fig:n_n_corr_Vm_25seed}(a) for $F$=0.5. All of them increase with $-V$ indeed. Note that after photoexcitation with $F$=0.6, the transient state is almost a negative-temperature state as suggested in Fig.~\ref{fig:op_F_U_Etot_25seed}(b). In negative-temperature states, the correlation functions generally behave as if the interactions were inverted.\cite{tsuji_prl11,tsuji_prb12,yonemitsu_jpsj15,yanagiya_jpsj15} Indeed, as shown in Fig.~\ref{fig:n_n_corr_Vm_25seed}(b) for $F$=0.6, the averaged $ \langle n_i n_j \rangle $ for any type of bonds decreases as $-V$ increases as if the attractive nearest-neighbor interaction $V$ were replaced by a repulsive one. 

\begin{figure}
\includegraphics[height=15.6cm]{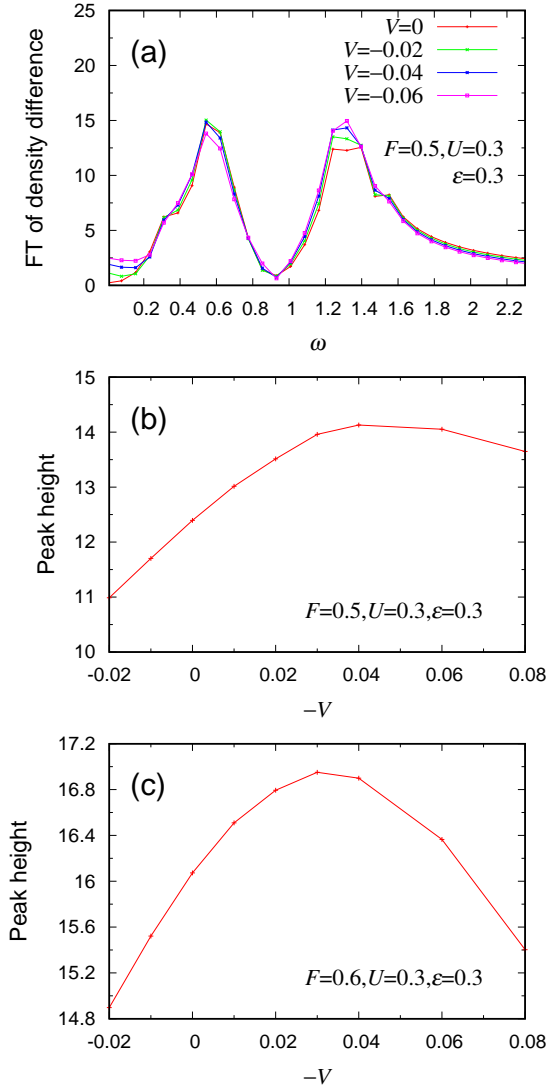}
\caption{(Color online) 
(a) Absolute values of Fourier transforms of random-number-distribution-averaged time profiles ($ T < t < 10T $) of charge-density difference between sublattices for different values of $V$, $F$=0.5, $U$=0.3, and $\epsilon$=0.3. (b) Peak height of Fourier spectrum (a) at $\omega \simeq$1.2 as a function of $-V$. (c) Peak height of Fourier spectrum at $\omega \simeq$1.2 as a function of $-V$ for $F$=0.6, $U$=0.3, and $\epsilon$=0.3. 
\label{fig:fourier_Vm_25seed}}
\end{figure}
\begin{figure}
\includegraphics[height=11.6cm]{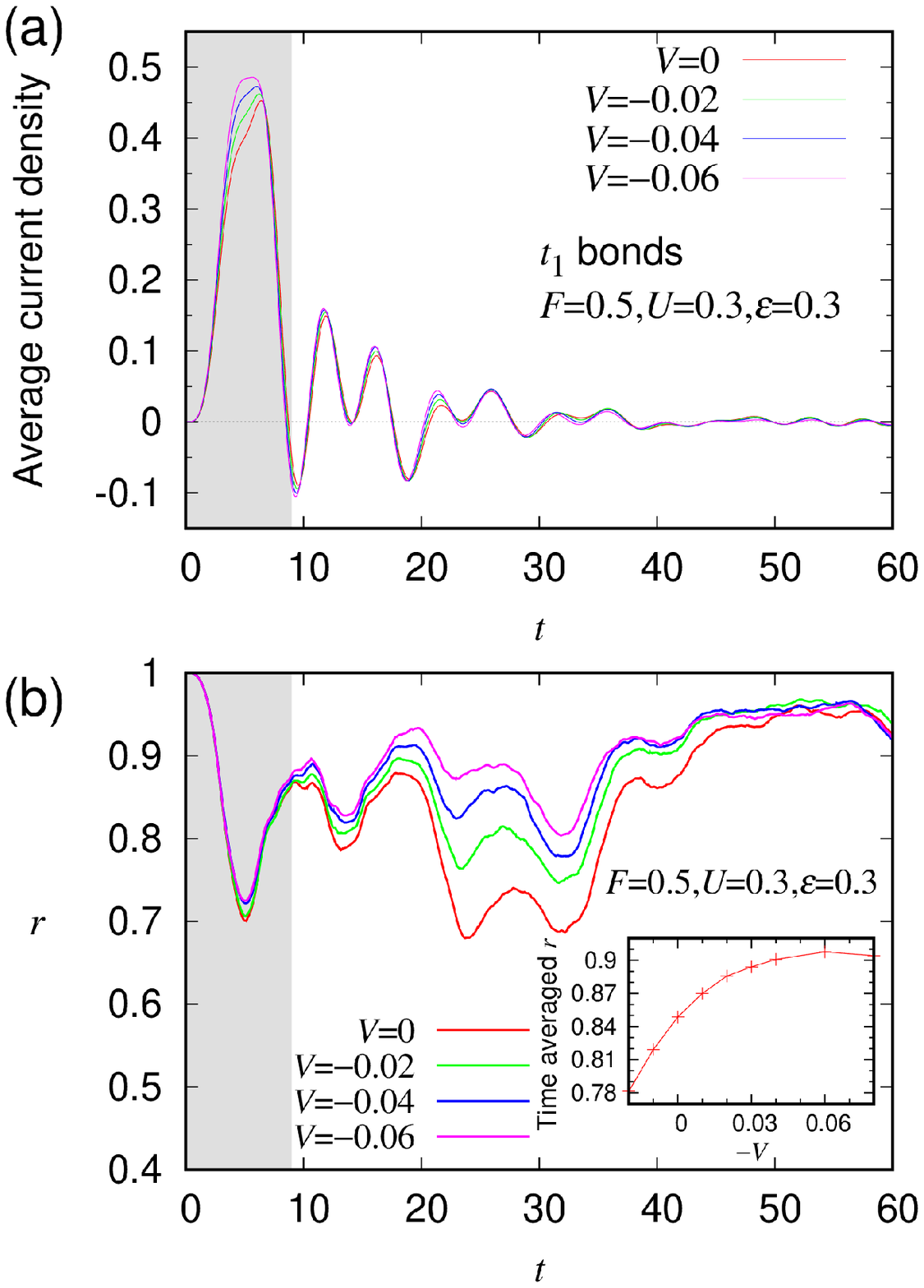}
\caption{(Color online) 
Random-number-distribution-averaged time profiles of (a) current densities on $t_1$ bonds and (b) synchronization order parameters for different values of $V$, $F$=0.5, $U$=0.3, and $\epsilon$=0.3. The inset of (b) shows the time average of the synchronization order parameter over $ 3T<t<6T $ as a function of $-V$ for the same parameters. 
\label{fig:cur_op_time_Vm_25seed}}
\end{figure}

If the magnitude of $V$ is large, it enhances dephasing and makes charge oscillations decay faster. If the magnitude of $V$ is small, a negative $V$ contributes to an increase in the magnitude of the current density on each bond, as shown later in Fig.~\ref{fig:cur_op_time_Vm_25seed}(a). Thus, a weakly attractive nearest-neighbor interaction $V$ increases the amplitudes of charge oscillations. This fact is consistent with the behavior shown in Fig.~\ref{fig:fourier_Vm_25seed}(a), which reveals that the Fourier spectrum of the charge-density difference averaged over random number distributions is heightened by the weakly attractive nearest-neighbor interaction $V$. The peak due to the electronic breathing mode at $ \omega_{\mbox{osc}} \equiv 2\left( \mid t_1 \mid + \mid t_2 \mid +2 \mid t_y \mid \right) $=1.2 is heightened and then lowered by increasing $-V$, as clearly shown in Figs.~\ref{fig:fourier_Vm_25seed}(b) and \ref{fig:fourier_Vm_25seed}(c). The charge oscillations are enhanced by the attractive interaction $V$ irrespective of whether the optical field amplitude $F$ is sufficiently large to produce a negative-temperature state [Fig.~\ref{fig:fourier_Vm_25seed}(c)] or not [Fig.~\ref{fig:fourier_Vm_25seed}(b)]. 

Next, we see how $V$ affects the charge oscillation dynamics. The time profiles of the current densities averaged over random number distributions and $t_1$ bonds are shown in Fig.~\ref{fig:cur_op_time_Vm_25seed}(a) for different values of $V$. Those on the other types of bonds show similar behaviors to those on $t_1$ bonds shown here. The duration of photoexcitation is shaded in this figure. In most of the time even after photoexcitation, a negative $V$ increases the magnitude of the current density on each bond if the magnitude of $V$ is small and its dephasing effect is thus small. In Fig.~\ref{fig:cur_op_time_Vm_25seed}(b), we show the time profiles of the synchronization order parameters averaged over random number distributions for different values of $V$. The order parameter substantially increases with $-V$ especially when it is small for $V$=0. Thus, the time-averaged order parameter increases with $-V$ more sensitively than the average current density for $-V\leq0.06$, as shown in the inset of Fig.~\ref{fig:cur_op_time_Vm_25seed}(b). 

\begin{figure}
\includegraphics[height=6.2cm]{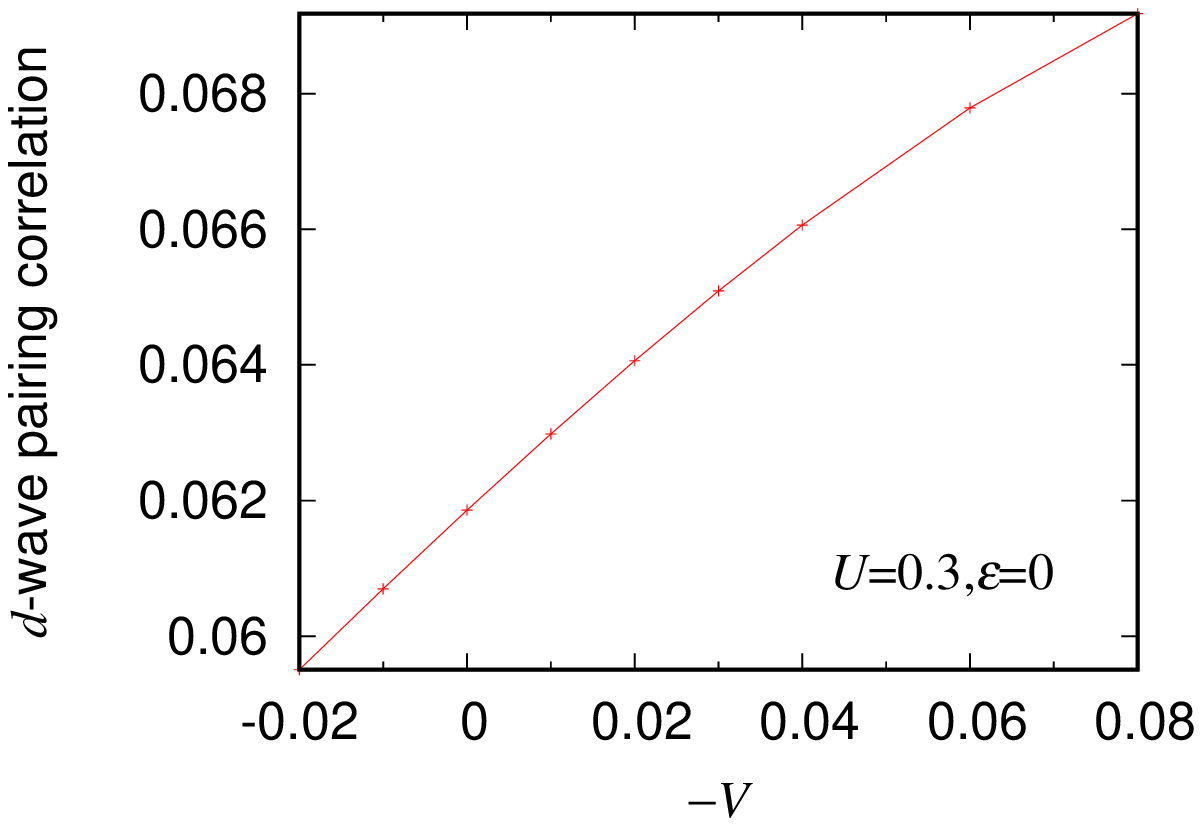}
\caption{(Color online) 
$d$-wave pairing correlation function $\langle \Delta_{\mbox{\boldmath $i$}} \Delta_{\mbox{\boldmath $i$+$r$}}^\dagger \rangle$ between farthest points with {\boldmath $r$}=(2$a$,2$a$) in ground state as a function of $-V$ for $U$=0.3 and $\epsilon$=0. 
\label{fig:d-wave_pairing_correlation_vs_vm_u0p3vmx_eps0}}
\end{figure}
This finding with respect to $V$ is consistent with recent experimental results. Although the Coulomb interaction in real systems is repulsive ($V>0$), the stimulated emission, which is supposed to be due to the electronic breathing mode,\cite{yonemitsu_jpsj18a} is enhanced by superconducting fluctuations in $\kappa$-(BEDT-TTF)$_2$Cu[N(CN)$_2$]Br.\cite{kawakami_np18} In this organic superconductor, the superconducting gap is considered to have the $d$-wave symmetry.\cite{kawamoto_prl95,mayaffre_prl95,carrington_prl99} Theoretically, in models with on-site repulsion $U>0$ and nearest-neighbor attraction $V<0$, a $d$-wave pairing correlation is generally enhanced\cite{micnas_prb88,micnas_prb89,micnas_rmp90,tsuchiura_jpsj95,murakami_jpsj00} unless phase separation is realized.\cite{w_p_su_prb04} 
Here, we consider a pairing of the form 
$ \Delta_{\mbox{\boldmath $i$}}^\dagger = 
c_{\mbox{\boldmath $i$}\uparrow}^\dagger 
c_{\mbox{\boldmath $i$}+\mbox{\boldmath $x$}\downarrow}^\dagger - 
c_{\mbox{\boldmath $i$}\uparrow}^\dagger 
c_{\mbox{\boldmath $i$}+\mbox{\boldmath $y$}\downarrow}^\dagger + 
c_{\mbox{\boldmath $i$}+\mbox{\boldmath $x$}\uparrow}^\dagger 
c_{\mbox{\boldmath $i$}\downarrow}^\dagger - 
c_{\mbox{\boldmath $i$}+\mbox{\boldmath $y$}\uparrow}^\dagger 
c_{\mbox{\boldmath $i$}\downarrow}^\dagger $, where bonds $\langle \mbox{\boldmath $i$},\mbox{\boldmath $i$}+\mbox{\boldmath $x$} \rangle$ and $\langle \mbox{\boldmath $i$},\mbox{\boldmath $i$}+\mbox{\boldmath $y$} \rangle$ are interdimer $t_2$ and $t_y$ bonds, respectively. We calculate $\langle \Delta_{\mbox{\boldmath $i$}} \Delta_{\mbox{\boldmath $i$+$r$}}^\dagger \rangle$ with {\boldmath $r$}=(2$a$,2$a$) in the ground state and show it in Fig.~\ref{fig:d-wave_pairing_correlation_vs_vm_u0p3vmx_eps0}, as a function of $-V$. It is indeed enhanced by the nearest-neighbor attraction. 
Although the realistic nearest-neighbor interaction is repulsive, the effect of increasing $d$-wave superconducting fluctuations with decreasing temperature in the experiment might be simulated to some extent by the attractive nearest-neighbor interaction in the present calculations. We suspect that superconducting fluctuations enhance an electron's transfer correlated with another electron's transfer, which would facilitate the synchronized charge oscillations and consequently the stimulated emission. However, the present system is too small to judge even whether it is metallic or insulating, which prevents quantitative discussions. 

\section{Conclusions}
On the basis of the previously reported emergence of an electronic breathing mode and synchronization of charge oscillations after strong photoexcitation on dimer lattices,\cite{yonemitsu_jpsj18a,kawakami_np18} we theoretically study the competition between the effect of interactions $ U $ and $ V $ and that of randomness $ \epsilon $ introduced into transfer integrals in an extended Hubbard model at three-quarter filling on a simple dimer lattice. For the definition of a synchronization order parameter, we use only current densities on bonds, derive phases $ \phi $ from their time profiles, and average $ e^{i\phi} $ over all bonds: it is defined only when current densities change in time. Owing to the randomness, current densities for $U$=$V$=0 on different bonds oscillate with different phases, so that the synchronization order parameter is small. 

When the optical field amplitude $F$ is large (but not too large to raise the entropy significantly), the on-site repulsion $U$ assists the charge oscillations to be synchronized and increases the order parameter. A sufficiently strong interaction $U$ overcomes the effect of randomness; thus, the order parameter almost reaches the maximum value. An even larger $U$ makes the charge oscillations decay faster through dephasing, so that it becomes difficult to observe the synchronization. As to the nearest-neighbor interaction $V$, a weakly attractive one enhances the synchronization by enhancing current flows. It is reminiscent of enhanced stimulated emission above the superconducting transition temperature where superconducting fluctuations are expected to assist it,\cite{kawakami_np18} in view of the fact that the stimulated emission is caused by an electronic breathing mode.\cite{yonemitsu_jpsj18a} However, the interaction $V$ is repulsive in real materials, and a repulsive interaction $V$ in small systems that can be treated by the exact diagonalization method decreases the synchronization order parameter, which is consistent with the previous result.\cite{yonemitsu_jpsj18a} The effect of superconducting fluctuations is beyond the scope of this study and left for future studies. 

\begin{acknowledgments}
This work was supported by JSPS KAKENHI Grant No. JP16K05459,                          MEXT Q-LEAP Grant No. JPMXS0118067426, and JST CREST Grant No. JPMJCR1901. 
\end{acknowledgments}

\bibliography{69713}

\end{document}